\documentclass[superscriptaddress,prd,aps,preprint,eqsecnum,floats]{revtex4}
\usepackage{epsfig}
\begin{document}
\title{Light meson radial Regge trajectories}
\author{A.M. Badalian}
\affiliation{State Research Center,\\ Institute of Theoretical and Experimental
Physics, Moscow, Russia}
\author{B.L.G. Bakker}
\affiliation{Department of Physics and Astronomy, Vrije Universiteit, Amsterdam}
\author{Yu.A. Simonov}
\affiliation{Jefferson Laboratory, Newport News, VA 23606, USA}
\affiliation{State Research Center,\\ Institute of Theoretical and Experimental
Physics, Moscow, Russia}
\begin{abstract}
A new physical mechanism is suggested to explain the universal
depletion of high meson excitations. It takes into account the
appearance of holes inside the string world sheet due to $q\bar{q}$
pair creation when the length of the string exceeds the critical value
$R_1 \simeq 1.4$ fm. It is argued that a delicate balance between large
$N_c$ loop suppression and a favorable gain in the action, produced by
holes, creates a new metastable (predecay) stage with a renormalized
string tension which now depends on the separation $r$.  This results in
smaller values of the slope of the radial Regge trajectories, in good
agreement with the analysis of experimental data in
Refs.~\cite{ref.3}.
\end{abstract} 
\maketitle
\section{Introduction} 
\label{sec.1}

Recently a number of radially excited mesons have been experimentally 
observed \cite{ref.1}-\cite{ref.4} and it was discovered that for 
excitations like $a_J(2P)$, $\omega_3(2D)$, and $\rho(2D)$ their masses
are 100-200 MeV lower than the theoretical predictions in
different models, in particular in the relativized potential model
(RPM)\cite{ref.5} and in the flux tube model \cite{ref.6}. A $K$-matrix
analysis of the Crystal Barrel data has shown that the Regge trajectories 
as a function of the radial quantum number $n_r$ continue to be
linear up to high excitations like the four $L$ states ($n_r=0,1,2,3$) and
can be described by the  $n_r$-trajectory (the radial Regge trajectory)
\cite{ref.3}:
\begin{equation}
 M^2(n_r,L)  = M^2(0,L) + \Omega_L\, n_r \quad ({\rm fixed} L)
\label{eq.1}
\end{equation}
where the slope $\Omega_L$ was found to vary in the narrow range $1.15
\leq \Omega \leq 1.30$ GeV${}^2$ for different $L$-wave states. 

In Ref.~\cite{ref.7} the orbital excitations of the light mesons
$(n_r=0,\, L \leq 5)$ were studied in detail in the framework of the
QCD string approach\cite{ref.8} and the spin-averaged meson masses
$\bar{M}(nL)$, the Regge slope $\alpha'_L$, and the intercept
$\alpha_L(0)$ were calculated analytically and expressed through a
single parameter, the string tension $\sigma$, while the Regge
intercept $\alpha_L(0)$ does not depend on $\sigma$ and is a universal
number. The calculated values of $\bar{M}(nL)$, $\alpha'_L$, and
$\alpha_L(0)$ turn out to be in very good agreement with the
experimental data.  (Note our notation: a state denoted by $nL$ has
radial quantum number $n_r = n-1$, so the lowest state with a given
angular momentum $L$ is $1L$ and has $n_r = 0$.)

The situation appears to be different for the radial excitations
(called also {\em radials} in what follows) calculated also for the
linear string potential (or the linear plus Coulomb potential) with the
same string tension.  Thus the masses of the radials obtained do lie on
a linear trajectory Eq.~(\ref{eq.1}) which has a slope that is
practically independent of $L$, but its value $\Omega_0 \simeq 2.0$
GeV${}^2$ appears to be a factor 1.6 - 1.5 larger than $\Omega$(exp),
the value extracted from the experimental data \cite{ref.3} (see
Sect.~2). In particular, the masses of the second radial excitations
(and even some first ones like $2D$ and $2F$) are 100-150 MeV $(n_r=2)$
(50 - 100 MeV for $n_r=1$) higher than $M$(exp) from Refs.~\cite{ref.1}
- \cite{ref.4}.

This phenomenon, the lowering of the masses of the radials, was already
discussed in Refs.~\cite{ref.9} and \cite{ref.10} where it was supposed
that this effect is connected with the opening of new channels, i.e.
with hadronic shifts. However, hadronic shifts cannot produce the global,
quantum-number independent shift down of all radials on the
$n_r$-trajectory having been observed in experiment. In particular it
cannot provide almost the same slope $\Omega$ in Eq.~(\ref{eq.1}) for
different $L$, since hadronic shifts strongly depend on the quantum
numbers of a decay channel, the closeness to the decay thresholds, the
widths, and many other specific features of meson decays
\cite{ref.11}.

Thus a basic paradox of mesonic spectra is that there are highly
excited meson states with large width, implying strong coupling to decay
channels, which nevertheless lie on linear Regge trajectories. This
situation implies that first, the string between a quark and an antiquark
continues to exist up to large energy excitations and can be as large
as $2.5-2.8$ fm (see below). Secondly, quark pair creation does not dominate
for such excitations, in particular the probability of string breaking is
not large.  How to reconcile these conclusions with strong decays
(large width) of the Regge-string mesons? One can argue that pair
creation is suppressed as $1/N_c$ at large $N_c$. Moreover, in
experiment this parameter appears to be 1/10 rather than 1/3 for
$N_c=3$ which can be seen in the width to mass ratio for large
excitations.  The present paper suggests at least a partial answer to
these questions taking as a characteristic example the radially excited
mesons and generalizing to all highly excited light mesons which serve
as a good illustration to the paradox stated above.  

So, to explain the ``global correlated shift down'' of the radial
excitations  we suggest here an alternative physical picture, which in
first turn takes into account the behavior of the string in highly
excited hadrons, and in addition the specific character of the
$\pi$-meson interaction with a light quark (antiquark) and the string
connecting them.

In contrast to hadronic decays like $\rho-\rho$, $\rho-a_0$,
$\omega-\omega$ etc., which may occur due to $q\bar{q}$ loop creation
inside the string's world sheet and subsequent string breaking, the
$\pi$-meson, as well as other Goldstone particles, locally interacts with
a quark (antiquark) sitting at the end of the string \cite{ref.12} and
therefore the string may not break due to the emission of a $\pi$-meson
from a quark.

The same statement is true when the creation of a $\pi$-meson is
accompanied by a $\rho$, $a_0$ or $f_0$ meson, since these mesons are
actually described by the remaining string  in the final state.
Therefore it is natural to assume that below the threshold of
$\rho-\rho$, ($\omega-\omega$ etc.), i.e. 
\begin{equation}
 E_{\rm thr} \simeq 2M_\rho \pm \Gamma_\rho  \geq 1.4 \,{\rm GeV}
\label{eq.2}
\end{equation}
the string effectively stays intact.  When only channels like
$(n\pi)\pi$, $(n\pi)\eta$, $\pi-\rho$, $\eta-\rho$ etc. are open, we
argue that the string between a light quark and antiquark is not
modified by open channels and has the same string tension
$\sigma_0\simeq 0.18$ GeV${}^2$ as the string between a static quark
$Q$ and antiquark $\bar{Q}$.  Then one can determine the characteristic
size of the string $R_1$ which corresponds to the value $E_{\rm thr}
\sim 1.4$ GeV:  
\begin{equation}
 \sigma_0 R_1  \simeq E_{\rm thr} ,
 {\rm or}\, R_1 \ge 1.4 \,{\rm GeV}/\sigma_0 = 1.45 \,{\rm fm}.  
\label{eq.3} 
\end{equation}

Thus our first assumption here is that up to distances $\simeq 1.4$ fm the
string potential is not distorted by meson decays. This statement is
in agreement with lattice calculations where in the presence of dynamical
fermions the static potential appears to be the same as in the quenched
approximation up to the separations of order $1.2-1.5$ fm \cite{ref.13},
\cite{ref.14}.

For higher excitations, $E^* \geq E_{\rm thr}$, the $q\bar{q}$ pair
creation inside the world sheet (with the quantum numbers ${}^3 S_1$ and
${}^3 P_0$) is already possible and at first sight the problem is
becoming essentially a two- or many-channel problem. However, we shall
assume and argue here (see Section \ref{sec.3}) that up to very high
excitations, $E_{\rm cr} \geq$ 2.5 GeV, i.e. in the range
\begin{equation}
 1.4 \; {\rm GeV} \leq  E^* \leq E_{\rm cr} \sim 2.5 \; {\rm GeV}
\label{eq.4}
\end{equation}
and for the time extensions $\delta T \sim 1/ \Gamma \leq 1.0$ fm, only
virtual loops or loops of small sizes are created. As a result the
probability of hadronic decays like $M \rightarrow \rho-\rho$,
$\omega-\omega$, is small while with a large probability the string
remains unbroken. This assumption is necessary to explain the linear
character of the $n_r$-trajectories Eq.~(\ref{eq.1}) up to high
excitations of the order of 2.5 GeV.

However, in the presence of such virtual or small $q\bar{q}$ loops the
string tension is renormalized and becomes dependent on the separation
$r$.  In the suggested picture (when at small $r$, $r \leq R_1$,
$\sigma= constant= \sigma_0$) the attenuation of the string tension is
being felt only at distances $r \geq R_1$ and continues at least up to
the value, $R_2 \sim2.5$ fm with $\sigma(R_2)= \sigma_2$. It is
important that this ``asymptotic" value $\sigma_2$ strongly affects the
slope $\Omega$ of the $n_r$-trajectory given in Eq.~(\ref{eq.1}).

The most important feature of this picture is the existence of a
prehadronization stage in the range of string parameters,
\begin{equation}
 R_1 \sim 1.4\; {\rm fm} \leq r \leq R_2 \sim 2.5 \; {\rm fm},
 \quad T \leq T_{cr}  \sim 1.0 \;{\rm fm},
\label{eq.5}
\end{equation}
where the string tension depends on $r$ while at the same time the
string with large probability remains unbroken. The dynamics in this
prehadronization region can be effectively described in the one-channel
approximation taking into account virtual quark loops and the open
hadronic channels (mostly like $(n\pi)\pi$, $\pi-\rho$, $\eta-\rho$
etc.) through a universal dependence of the string tension on $r$:
\begin{eqnarray}
 \sigma  & = & constant = \sigma_0, \quad   r \leq R_1, \nonumber \\
 \sigma  & = & \sigma(r)  \rule{23mm}{0mm}R_1 \leq  r \leq R_2
\label{eq.6}
\end{eqnarray}
At the present stage of the theory the function $\sigma(r)$ is not yet
calculated in full QCD and therefore we formulate here the problem in a
different way: how to extract information about $\sigma(r)$, or the
string breaking, from highly excited meson masses, in particular from
the slope of the radial Regge trajectories.

We shall show here that there exists a direct connection between the
slope $\Omega$ and the two most important features of $\sigma(r)$: the
value of $R_1$ where the string tension is becoming $r$-dependent, and
the value $\sigma_2$ which characterizes the string tension in the
region where breaking is already possible.  In lattice calculations a
flattening of the  static potential due to $q\bar{q}$ pair creation at
$r \sim 1.2-1.5$ fm seems to be observed \cite{ref.14}, unfortunately,
lattice points have very large errors and at present definite
conclusions about the exact value and form of the static
potential at large $r$ cannot be derived from lattice measurements.

We concentrate below on these considerations and suggest a workable and
simple model for the mesons of large size, both radial and orbital
ones, which yields meson masses in good agreement with experiment. It
will be shown that in the proposed picture the linear
$n_r$-trajectories with a rather small slope, $\Omega \simeq 1.3$
GeV${}^2$ ($\simeq 1.5$ GeV${}^2$ for the spin averaged $S$-wave
states) close to the experimental numbers, can be easily obtained.

The plan of the paper is as follows. In Section~\ref{sec.2} the
analytic formulas for the Regge slope and the intercept for the linear
potential are derived and the masses will be expressed through a single
scale parameter - the string tension. It will be shown that for the
standard linear potential the slope $\Omega_0$ is a factor 1.6 larger
than in experiment. In Section~\ref{sec.3} the effects of
$q\bar{q}$-pair creation (unquenched situation) on the meson masses of
large radii are discussed and in Section~\ref{sec.4} a modified
nonperturbative potential is proposed for which the meson masses are
calculated. In Section~\ref{sec.5} the Regge slope and intercept of the
$n_r$-trajectories are presented. In section~\ref{sec.6} our
Conclusions and some prospectives are briefly discussed. In the
Appendix the results of the detailed calculations for the meson spectra
are included.

\section{The problem of high excitations for the linear potential}
\label{sec.2}

In the QCD string approach the Hamiltonian is derived from QCD under
definite and verifyable assumptions. First, the times and distances
involved are considered to be larger than the gluonic correlation
length $T_g$: $r \gg T_g$ ($T_g\simeq 0.2$ fm in lattice calculations
\cite{ref.15}). This condition is always valid for the light mesons
having large sizes $R \geq 0.8$ fm. Secondly, the string (hybrid)
excitation scale is large, $\Delta M_{\rm str} \geq 1$ GeV, and
therefore in first approximation the meson and hybrid excitations are
disconnected.

Then the spin-averaged mass $\bar{M}(nL)$ of the light meson with
arbitrary quantum numbers $nL$ ($L \leq 5$)are determined by the
following mass formula \cite{ref.7,ref.8}:
\begin{equation}
 \bar{M}(nL) = M_0(nL) + \Delta_{\rm str}(nL) + \Delta_{\rm SE}(nL)
\label{eq.8}
\end{equation}
where $M_0$ is the eigenvalue of the unperturbed string Hamiltonian $H_R^1$ 
coinciding with the spinless Salpeter equation (SSE):
\begin{equation}
 \left(2\sqrt{\vec{p}^2 + m^2}  + V(r) \right) \psi(nL) = M_0(nL)\psi(nL),
\label{eq.9}
\end{equation}
where $m$ is the current quark mass taken here to be equal to zero
($m=0$).  The potential $V(r)$ contains in the general case both a
perturbative part, the Coulomb interaction $V_C(r)$, and the
nonperturbative string potential $V_{\rm NP}(r)$ with a string tension
that is in general dependent on the separation $r$ \cite{ref.7}:
\begin{equation}
 V(r) = -\frac{4\alpha_s}{3r} + \sigma(r) r .
\label{eq.10}
\end{equation}
It is instructive to consider first a linear potential with
$\sigma=constant = \sigma_0$ with a mass formula that is more transparent
and can be presented in analytical form. In this case the string
correction was defined in \cite{ref.7,ref.16}:
\begin{eqnarray}
 \Delta_{\rm str}(nL)= - \frac{\sigma_0<r^{-1}>L(L+1)}{8\mu_0^2(nL)} 
 = - \frac{2\sigma_0<r^{-1}>L(L+1)}{M_0^2(nL)} ,
\label{eq.11}
\end{eqnarray}
where the following relations valid for the potential $\sigma_0 r$ are used:
\begin{equation}
 M_0(nL) = 4 \mu_0(nL), \langle \sigma_0\, r \rangle = 2 \mu_0(nL) .
\label{eq.12}
\end{equation}
The constituent mass $\mu_0(nL)$ was derived to be the average of the
quark kinetic energy operator \cite{ref.7},\cite{ref.8}:
\begin{equation}
 \mu_0(nL)  = \langle \sqrt{\vec{p}^2 + m^2}\rangle_{nL}.
\label{eq.13}
\end{equation}
From the definition Eq.~(\ref{eq.13}) it is clear that $\mu_0$ depends
on the quantum numbers $nL$ of a given state and can be expected to grow
for high excitations.

An important contribution to the meson mass Eq.~(\ref{eq.8}) comes from the
nonperturbative self-energy term $\Delta_{\rm SE}(nL)$ calculated in
Ref.~\cite{ref.17}:
\begin{equation}
 \Delta_{\rm SE}(nL) = - \frac{4\sigma_0\eta(f)}{\pi\mu_0}
		     = - \frac{16\sigma_0\eta(f)}{\pi M_0(nL)},
\label{eq.14}
\end{equation}
where the parameter $\eta(f)$ depends on the quark flavor $f$ and can
be calculated, see Ref.~\cite{ref.17}:
\begin{equation}
 \eta(n\bar{n})  = 0.90 
\label{eq.15}
\end{equation}
Here it is worth to notice that the Coulomb corrections $E_C(nL)$ to
the light meson masses are small, $|E_C| \leq 100$ MeV and can be
neglected in the first approximation.  The results of the exact
calculations for a linear plus Coulomb potential Eq.~(\ref{eq.10}) will
be presented in Section~\ref{sec.4} and in the Appendix.

The self-energy term enters the squared mass $\bar{M}^2$ in such a way that
it gives rise to an important negative constant $C_0$,
\begin{equation}
 C_0 = - \frac{32\eta\sigma_0}{\pi} 
\label{eq.16}
\end{equation}
so that the mass formula for the squared meson mass is
\begin{equation}
 \bar{M}^2(nL)= M_0^2 - \frac{4\sigma_0<r^{-1}>L(L+1)}{M_0}
 - \frac{32\eta\sigma_0}{\pi} + \frac{256\sigma^2_0\eta^2}{\pi^2M_0^2},
\label{eq.17}
\end{equation}
i.e. it is proportional to $\sigma$ and does not contain any free
parameter. In Eq.~(\ref{eq.17}) the contributions of the small terms
$\Delta_{\rm str}^2$ and $2\Delta_{\rm str}\Delta_{\rm SE}$ were
neglected to show the most important features of the meson spectra. On
the contrary the term $\Delta_{\rm SE}^2 \geq 0.10$ GeV${}^2$ is not
small for any state and therefore is kept in the mass formula
(\ref{eq.17}).

An important next step refers to the approximation for the eigenvalues
$M_0^2 (nL)$ of the SSE valid for the linear potential $\sigma_0r$ :
\begin{equation}
 M_0^2({\rm approx})= [8L + 4\pi\xi(nL)\,n_r + 3\pi]\sigma_0.
\label{eq.18}
\end{equation}
This formula with $\xi=1.0$ reproduces the exact values of $M_0^2(nL)$
with an accuracy better than 2\% for all $nL$-states with the exception
of the $1S$ and $1P$ states where the accuracy is 3-6\% \cite{ref.7}.
In Eq.~(\ref{eq.18}) the coefficient $\xi(nL)\simeq 1.0$ weakly depends
on $L$ and slightly decreases with growing $n_r$, e.g. for the $4S$ state
$\xi(4S) = 0.99$ while $\xi(4F) = 0.96$. In what follows in most cases
we put $\xi(nL) = 1.0$.

Then with the use of the expression (\ref{eq.18}) and redefining the
matrix element $<r^{-1}> = \sqrt{\sigma_0} <\rho^{-1}>$ where
$<\rho^{-1}>$ is already independent of $\sigma_0$, one can rewrite the
mass formula Eq.~(\ref{eq.17}) in the following way,
\begin{equation}
 \bar{M}^2(nL)= (\alpha_L')^{-1}\,L + [4 \pi n_r +b(nL)]\sigma_0
\label{eq.19}
\end{equation}
giving for the $L$-trajectory,
\begin{equation}
  L = \alpha'_L\;\bar{M}^2(nL) + \alpha_L(n), 
\label{eq.20}
\end{equation}
the Regge slope (in general $n_r \neq 0$):
\begin{equation}
 (\alpha'_L)^{-1} = \sigma_0 [ 8 - \delta(nL)] \equiv
 \sigma_0 \left[8 - \frac{\surd 2 <\rho^{-1}>(L+1)}
  {\sqrt{L+ \frac{\pi n_r}{2} + \frac{3\pi}{8}}} \right].
\label{eq.21}
\end{equation}
For the leading $L$-trajectory ($<\rho^{-1}> \sqrt{L+1} \sim $ constant)
the slope is given by the constant \cite{ref.7}
\[
 (\alpha'_L)^{-1} = (6.95\pm0.02) \sigma_0 
\]
or
\begin{equation}
 \alpha'_L = \frac{1}{6.95\sigma_0} = 0.80 \, {\rm GeV}^{-2}
 \quad {\rm for } \quad \sigma_0=0.18 \, {\rm GeV}^2,
\label{eq.22}
\end{equation}
which is in good agreement with the experimental value $\alpha'_L({\rm
exp}) = 0.81 \pm 0.02$ GeV$^{-2}$. For the orbital excitations with a
fixed $n_r >0$, $\alpha_L'(n_r)$ appears to be a bit smaller since for
them $\delta(nL)$ in Eq.~(\ref{eq.21}) is smaller, e.g.
$\delta(n_r=1)=0.85 \pm 0.05$ and $\delta(n_r=3) \simeq 0.50$ while 
$\delta(n_r=0)=1.05$. Then for $\sigma_0= 0.18$ GeV${}^2$
\begin{eqnarray}
 \alpha_L'(n_r=1) & = & 1/[7.2\sigma_0] = 0.77 {\rm GeV}^{-2}, \nonumber \\
 \alpha_L'(n_r=3) & = & 1/[7.5\sigma_0] = 0.74 {\rm GeV}^{-2}
\label{eq.23}
\end{eqnarray}
However, this slope gives larger values for the radial excitations, e.g.
$\bar{M}(2P)=1.82\pm0.03$ GeV while the expected experimental number is
$\bar{M}(2P)\leq 1.70- 1.75$ GeV \cite{ref.2}, \cite{ref.3}.

From Eq.~(\ref{eq.17}) it follows that the constant $b(nL)$ in
Eq.~(\ref{eq.19}) is
\begin{equation}
  b(nL)= 3\pi - \frac{32\eta}{\pi} +\frac{32\eta^2}
 {\pi^2 (L+ \frac{\pi n_r}{2}+ \frac{3 \pi}{8})}.
\label{eq.24}
\end{equation}
As was stressed in Ref.~\cite{ref.7}, for the $b(1S)$ meson it is more
precise to use the exact eigenvalue with  $M_0^2 (1S)= 9.967 \,
\sigma_0$ instead of the approximation (\ref{eq.18}) so that for the
leading $L$-trajectory ($\eta = 0.90$) one finds
\begin{eqnarray}
 b(1S) & = & 3\pi - \frac{32\eta}{\pi} 
 +\frac{256\eta^2\sigma_0}{\pi^2M^2_0}, \nonumber\\
 & = & 0.257+\frac{256\eta^2}{9.967\,\pi^2}  =  2.365 .
\label{eq.25}
\end{eqnarray}
Then from the mass formula (\ref{eq.19}) the intercept of the leading
$L$-trajectory,
\begin{eqnarray}
 \alpha_L(\bar{M}^2=0)= - {\alpha'_L} b(1S)\,\sigma_0 = - 2.365/6.95 = - 0.34,
\label{eq.26}
\end{eqnarray}
does not depend on the string tension and is a universal number. With
10\% accuracy it coincides with the experimental number
$\alpha_L(0)_{\rm exp} = 0.30 \pm  0.02$ \cite{ref.7}.

For $n_r\neq 0$ the intercept of the $L$-trajectory  is
\begin{eqnarray}
 \alpha_L(n_r,\bar{M}^2 =0) & = & - \alpha_L'(n_r)[4\pi n_r + b(nL)]\sigma_0
 \nonumber \\
 & = & - \frac{0.257 + \frac{2.626}{L+\pi n_r/2 + 3\pi/8} + 4\pi n_r}
 {8-\delta(nL)}\quad (n_r \neq 0) .
\label{eq.27}
\end{eqnarray}
In the intercept (\ref{eq.27}) the term $b(nL)$ is small as compared
to $4 \pi n$ (the largest  correction  is for the $S$-wave states but
even then it is $\leq 10\% $) and can be neglected in first approximation.
Then also neglecting $\delta(nL)$ compared to 8 one obtains
\begin{equation}
 \alpha_L(n_r) \simeq  - \frac{\pi n_r}{2}, 
\label{eq.28}
\end{equation}
which does not depend on $L$ and  $\sigma_0$. For the neighbouring
intercepts the difference is equal to the following constant,
\begin{equation}
 \alpha_L(n_r+1)  - \alpha_L(n_r) = - \pi/2 = - 1.57.
\label{eq.29}
\end{equation}
The magnitude of the intercept is rather large due to the
presence of the large number $4\pi n_r\sigma_0$ in the eigenvalue
$M_0^2(nL)$ (or $\bar{M}^2$ (\ref{eq.19})). Note that another large
number, $3\pi \sigma_0$, which is present in $M_0^2(nL)$, is practically
cancelled by the constant $C_0$, Eq.~(\ref{eq.16}), coming from the
self-energy contribution.  Now one can present the masses of the
radials in the form of the $n_r$-trajectory Eq.~(\ref{eq.1}):

\begin{equation}
 \bar{M}^2(n_r,L) = M^2(0,L) + \Omega_0 \,n_r\quad   (L\, {\rm fixed}),
\label{eq.30}
\end{equation}
where from the mass formula (\ref{eq.17}) one finds for the linear potential 
the following expression for $\Omega_0$
\begin{equation}
 \Omega_0 = [4\pi\sigma_0  \xi(nL)] [1 - \chi (nL)]
 \simeq 1.90-2.10 \;{\rm GeV}^2 \quad
 {\rm for}\, \sigma_0=0.18\, {\rm GeV}^2.
\label{eq.31}
\end{equation}
($\chi(nL) = (b(nL) - b(n+1\,L))/4\pi$)

In Eq.~(\ref{eq.31}) $0.95 \leq \xi(nL) \leq 1.05$ and $0.05 \leq  \chi
\leq 0.10$ for all $nL$ states ($n_r \leq4, L \leq 4$) so that
$\Omega_0$ is only 5\%-10\% smaller than $4\pi\sigma_0$. 

Thus for the linear potential we have obtained\\
(i) The masses of the radials lie on linear $n_r$-trajectories as in 
Eq.~(\ref{eq.1});\\
(ii) The slope $\Omega _0$ does not depend on $L$ (with 95\% accuracy) in
agreement with experimental observations;\\
(iii) However, numerically the value of $\Omega_0$ turns out to be 
$\simeq 1.6$ times larger than the value extracted from the experimental 
data \cite{ref.3}.

To explain this phenomenon we suggest below a physical mechanism which
can be applied to highly excited mesons.

\section{Meson masses and quark pair creation} 
\label{sec.3}

The effective Hamiltonian derived from the QCD Lagrangian with the use
of the Fock-Feynman-Schwinger representation\cite{ref.18} is based on
the quenched approximation, where the quark determinant is replaced by
unity. Based on the width-to-mass ratio and on the existence of linear
Regge trajectories for the mesons it is usually argued that the effects
of the sea quark loops coming from the quark determinant cannot be
large and are estimated to be around 10\%. The same estimate of this
correction is obtained from lattice calculations for the unquenched
low-lying hadrons\cite{ref.19}. Moreover, lattice calculations of the
$Q\bar{Q}$ static potential up to $1.0-1.5$ fm do not show a significant
difference between quenched and unquenched calculations
\cite{ref.13}-\cite{ref.20}.

It can be shown that the radial excitations and high orbital
excitations ($L \geq 4$) have sizes exceeding 1.5 fm (see
Table~\ref{tab.1}) and therefore one should reconsider possible effects
of quark loops on the large-size mesons. A dedicated study on the
lattice \cite{ref.14} shows a flattening of the static potential or
decreasing of the string tension at separations $r > 1$ fm. However, at
present lattice points have large errors at such distances, quickly
deteriorating  with increasing $r$ and one cannot extract the exact
form of the static potential at large $r$ from lattice data.

Above the threshold of $q\bar{q}$ pair creation the static $Q\bar{Q}$
pair could decay into two heavy-light mesons with mass $M_{\rm HL} =
2m_Q + 2E_{\rm HL}$ where $E_{\rm HL}$ is the excitation energy of the
heavy-light meson which can be calculated in the framework of the
formalism presented in Ref.~\cite{ref.21} and is found to be $E_{\rm
HL}(\alpha_s=0)= 0.73$ GeV and $E_{\rm HL}(\alpha_s =0.39)= 0.53$ GeV
for a $b$-quark mass $m_b = 4.8$ GeV.

For the $c$ quark with  $m_c = 1.4$ GeV the values are close: $E_{\rm
HL}(\alpha_s = 0) = 0.76$ GeV and $E_{\rm HL}(\alpha_s =0.39) = 0.58$
GeV are obtained\cite{ref.22}. So one can expect that for lighter
quarks $E_{\rm LL}$ is also $\simeq 0.7-0.6$ GeV and the threshold of
$q\bar{q}$ pair creation inside the world-sheet of the string to be
$M_{\rm thr} = 1.2- 1.4 $ GeV in accord with the experimental $\rho-\rho$
threshold Eq.~(\ref{eq.2}).  Expressing this value in terms of the
distance $R_1$: $M_{\rm thr} = \sigma_0 R_1$  one finds the number
\[
	  R_1 =   1.3  - 1.5 \;{\rm fm}
\]  
close to the value $R_1$ in Eq.~(\ref{eq.3}).

At this point one should stress that the phenomenon discussed, viz the
pair creation just on the string, does not necessarily exhausts all
possible meson decay mechanisms or the true hadronization of the
mesons. Namely, as was shown in Ref.~\cite{ref.12}, pions are directly
coupled to the quark (antiquark) at the ends of the string and can be
emitted from there without breaking it.

Indeed, bosonization of quark degrees of freedom in Ref.~\cite{ref.12}
leads to the following term in the Lagrangian in the local limit:
\begin{equation}
 \Delta {\cal L}^{(1)} = \int dt  d^3 x \left[ \bar{q}(x)\sigma |\vec{x}|
 \gamma_5 \frac{\pi^a \lambda^a} { F_\pi} q(x) \right],
\label{eq.32}
\end{equation}
where the string starts from the antiquark position $\vec{x}=0$; the
field $\pi^a$ is the octet of Nambu-Goldstone mesons, and $F_\pi= 93$
MeV.

The same operator $\Delta {\cal L}^{(1)}$ between quark bound states
can be rewritten with the use of the Dirac equation as follows
\begin{equation}
 \Delta{\cal L}^{(1)} = g_A^q
 {\rm Tr}\,\bar{q}\gamma^\mu \gamma_5 \omega_\mu q ,
\label{eq.33}
\end{equation}
with
\begin{equation}
 \omega_\mu=\frac{i}{2F_\pi}\,
 ( u\partial_\mu u^\dagger - u^\dagger \partial_\mu u) \;
 u = \exp \left(i\gamma_5\frac{\pi^a \lambda^a} {2F_\pi} \right),
 \; g_A^q=1.0.
\label{eq.33a}
\end{equation}
From this expression one can see that $\Delta{\cal L}^{(1)}$ describes
the emission of an arbitrary number of Nambu-Goldstone mesons from the
quark position and therefore may describe pionic and double pionic
hadron decays while the string plays the role of a spectator and stays
intact.

At the same time $q\bar{q}$ pairs around the string (sea quarks) should be
identified with the loops of the determinant which can be written as
\cite{ref.22,ref.23}:
\begin{equation}
 \ln\det(m^2 + D^2)= \int d^4 x \int_{0}^{\infty}
 \frac{ds}{s}\,\exp(-m^2 s)
 (Dz)_{xx} \exp \left(-\int_{0}^{s}\frac{\dot{z}^2 d\tau}{4}\right)
 {W(C)} .
\label{eq.34}
\end{equation}
The integral $(Dz)_{xx}$ in Eq.~(\ref{eq.34}) is taken along the closed
loop $C$ from some point $x$ back to $x$ and contains an integral over
loops of all sizes. Alternatively, one can separate the quark
determinant into parts of small and large eigenvalues as it
Ref.~\cite{ref.14},
\begin{equation}
 \det(A) = \det{}_{\rm IR}(A) \det{}_{\rm UV}(A) ,
\label{eq.35}
\end{equation} 
where $\det_{\rm IR}$ takes into account the small eigenvalues (large
loops) $\lambda_n \leq \Lambda_{\rm cut}$ while $\det_{\rm UV}$
contains the large eigenvalues. The latter correspond to the
contribution of small virtual loops which should be properly renormalized.

In the physical picture $\det_{\rm IR}$ corresponds to the chiral
effects which disappear in $\det(m^2 + D^2)$ for large $m_q$:
$ m_q^2 \gg \Lambda_{\rm QCD}^2$ together with effects of large loops.

At this point it is important to stress that treating the resonances as
quasistationary states one should always consider Green's functions and
correspondingly Wilson $q\bar{q}$ loops of finite time extension T,
$T \sim 1/ \Gamma(nL) \sim 1 - 2$ fm.

For such finite times one can write a general expansion for the
original Wilson loop of the $q\bar{q}$ meson with pair creation due to
the quark determinant Eq.~(\ref{eq.34}) as follows
\begin{eqnarray}
 \langle W(C) \det(m+D) \rangle = W_0(C) +
 \frac{a_1}{N_c} W_1(C,C_1) + \frac{a_2}{N_c^2} W_2(C,C_1,C_2)  + \dots,
\label{eq.36}
\end{eqnarray}
where the coefficients $a_i =O(1)$ and higher Wilson loops $W_i, i\geq 1$,
are averaged products of the $i+1$ Wilson loops, i.e.
\begin{equation}
 W_1(C,C_1)= \langle\langle W(C)W(C_1) \rangle\rangle,\;
 W(C) = \frac{1}{N_c} {\rm Tr}\hat{W}(C) .
\label{eq.37}
\end{equation}
It is clear that when $T \to \infty$ and $r \geq R_1$ (so that 
decay is energetically possible) the asymptotics of the r.h.s. of
Eq.~(\ref{eq.36}) is
\begin{equation}
 \exp(-\sigma rT) + \frac{a_1}{N_c} \exp(-T (M_1+ M_2))
 + \frac{a_2}{N_c^2} \exp(-T \sum M_i) + \dots ,
\label{eq.38}
\end{equation}
where $\sum M_i$ is the sum of the masses of the decay products
($i=1,2$ in  the simplest case and  $i=1 \dots k$ for meson decay into
$k$-particles etc.). So, if $T \to \infty$ the resonances are dying out
and the second term in Eq.~(\ref{eq.38}) is dominant, which means that
only the products of meson decay are left.

However, for finite $T \simeq 1-2$ fm the situation is different and one can
expect a delicate balance between the large $N_c$ limit (suppressing
$q\bar{q}$ pair creation) and large times $T$ (preferring large internal
loops). This statement is the dynamical basis of our main assumption
about the existence of a specific state of the string with
effectively a hole inside the world sheet which obeys the area law with a
reduced (renormalized) string tension $\sigma^* = \sigma(r)$.

Thus we introduce the new concept of a transitional regime which in
terms of $Q\bar{Q}$ separations $r$ refers to the region
\[
 R_1 \leq r \leq R_2
\]
where $R_1 \sim 1.2 -1.4$ fm. $R_2\sim 2.5$ fm corresponds to high
excitations with energies $E^* \sim 2.5$ GeV.

In this region, due to relatively small quark loops and
correspondingly small holes in the string world sheet, the string
tension is nonperturbatively renormalized and decreases with growing
$r$. While the loops are still virtual, their presence does not lead to
actual string breaking (with appreciable probability) even though the
energy of the string for $r \ge R_1$ is sufficient for meson decay. We
suggest here that only for much larger distances, $r \geq R_2 \simeq
2.5$ fm the string breaking happens with large probability.

A first argument in favour of this picture is that highly excited mesons
of large radii do exist and lie on the corresponding linear Regge
trajectories, while their characteristics can be computed as in the QCD
string approach, neglecting decay channels. At the same time at such
large distances the string tension cannot remain intact and should be
strongly decreased by the appearance of holes inside the string world
sheet due to pair creation.

The second argument refers to the experimental information about strong
decay modes. The first strong decays without  Goldstone particles is
observed only for $f_0 (1370) \rightarrow \rho-\rho$ and $f_2(1565)
\rightarrow \rho_0-\rho_0$ decays in accord with our picture that
$\sigma=constant$ at distances $r \leq R_1$ or excitations $E^*\leq
1.4$ GeV.

For higher excitations, in the range 1.4 GeV $ \leq  E^* \leq 2.5$
GeV, at present only several strong decays like $f_2(1640) \rightarrow
\omega-\omega$, (seen); $\pi_2(1670) \rightarrow \omega-\rho$
(branching $\sim 2.7$\%), and $f_4 ( 2050) \rightarrow \omega-\omega$
(branching $\sim 25$\%) have been measured and in all these cases the
branching ratios of decays without Goldstone particles are never large.

The third argument refers to the r.m.s. radii $R(nL)= \sqrt{<r^2>_{nL}}$
of the radials which are ($n_r \neq 0$) calculated for the linear
potential with $\sigma_0 = 0.19$ GeV${}^2$ (see Table~\ref{tab.1}). (It
is worth to notice that the numbers given for $R(nL)$ represent the
lower limits of the true r.m.s. radii, since they correspond to larger
meson masses and the actual values of $R(nL)$ are about 20-50\% larger
(see Table~\ref{tab.6}).
\begin{table}
\caption{The light meson r.m.s. radii $R(nL)$ in fm for the $\sigma_0r$
potential with $\sigma_0 = 0.19$ GeV${}^2$.}
\label{tab.1}
\begin{tabular}{|l|r|r|r|r|r|}
\hline
 $n_r\setminus L$ &   0   &        1    &     2    &    3   &      4 \\
\hline
 0       & 0.80  &    1.02   &   1.21 &      1.37 &   \underline{1.51} \\
 1       & 1.22  & \underline{1.31} & \underline{1.52} & \underline{1.70} 
 & \underline{1.77} \\
 2       & \underline{1.53} & \underline{1.66} & \underline{1.78}
 & \underline{1.90} & \underline{2.00} \\
\hline
\end{tabular}
\end{table}
From the values of $R(nL)$ given here one can see that among the ground
states only the $1G$ state has $R(1G) \simeq 1.4$ fm, while for the $2L$
and the $3L$ states (with the exception of the $2S$ state with relatively
small r.m.s. radius equal to 1.21 fm) $R(nL) $are in the range:
\begin{eqnarray}
 1.4 \,{\rm fm}\, < R(2L) < 1.8 \,{\rm fm}\,  & (n_r=1),& \nonumber \\
 1.7 \,{\rm fm}\, < R(3L) < 2.0 \,{\rm fm}\,  & (n_r=2).&
\label{eq.39}
\end{eqnarray}
The values of $M$(exp) for all states underlined in Table~\ref{tab.1}
are shifted down compared to the theoretical values calculated with the
same linear potential that gives a good description of the orbital
excitations with $n_r=0$. This example agrees with our estimate of the
characteristic size $R_1 = 1.4$ fm where the pair creation is beginning
to affect the string tension $\sigma(r)$.

The r.m.s. radii in Table~\ref{tab.1} also show that even for the 
linear potential $R(nL) \geq 1.90$ fm for such states as the $3F$ states: 
$f_4(2290)$, $f_3(2280)$, $a_4(2280)$, and $a_3(2310)$\cite{ref.2}.

At this point it is important to stress the difference and similarity
of our approach with  that in Refs.~\cite{ref.9}, \cite{ref.10}. In both
approaches it is stressed that the $q\bar{q}$ pair creation is
responsible for the renormalization of the string tension and therefore
the unquenched string tension is lower than in the quenched case.
Moreover in Ref.~\cite{ref.10} as well as in the present paper it is
emphasized that even when level crossing occurs, i.e. when $V(r)$
equals $2M_{\rm HL}$, the string potential can be used throughout, only
the string tension being renormalized.

The difference between both approaches is in the meaning of this
renormalization. We assume here that there exists a universal (quantum
number independent) prehadronization stage when small quark loops
attenuate the string tension, while one can still neglect the influence
of specific decay channels which produce the hadronic shift.

It was already realized in Refs.~\cite{ref.9,ref.10} that the very fact
of the occurrence of smooth Regge trajectories and ordered hadronic
spectra is difficult to explain if the hadronic shifts are essentially
important, since the latter depend on the concrete hadronic channels
involved and vary irregularly from channel to channel.

In the suggested picture due to the universal predecay (prehadronization)
stage with renormalized (attenuated ) string tension not only can the spectrum
be calculated, but the notion of linear Regge trajectories, both
radial and orbital, is kept intact.

\section{Modified confining potential} 
\label{sec.4}

From the physical picture discussed in Section~\ref{sec.3} and the
introduction it follows that up to the characteristic distance $R_1
\sim 1.2 -1.4$ fm the string tension $\sigma$ is constant, while for
larger $r$ it depends on the $q\bar{q}$ separation as in
Eq.~(\ref{eq.6}). We propose here the nonperturbative potential
$\sigma(r)\,r$ with the string tension taken in the following form:
\begin{eqnarray}
 V_{\rm NP}(r) = \sigma(r)\,r, \quad
 \sigma(r) = \sigma_0 \left[1 - \gamma\frac{\exp(\sqrt\sigma_0 (r - R_1))}
 {B + \exp(\sqrt\sigma_0 (r - R_1))} \right].
\label{eq.40}
\end{eqnarray}
In the definition Eq.~(\ref{eq.40}) the constant $\gamma$ determines the
value of the string tension at large separations so that
\begin{equation}
 \sigma_2 =\sigma(r \geq 2.5 \,{\rm fm}) \simeq \sigma_0(1 - \gamma ).
\label{eq.41}
\end{equation}
Note that $\sigma_0 \sim 0.18 -0.19$ GeV${}^2$ defines the common scale
of the modified string potential and can be fixed by the Regge slope of
the leading $L$-trajectory.

In general our calculations are performed with the potential $V(r)$
Eq.~(\ref{eq.10}) containing the perturbative potential $V_{\rm P}(r)$,
\begin{equation}
  V_{\rm P}(r) = - \frac{4\alpha_s}{3r},
\label{eq.42}
\end{equation}
where for the strong coupling constant the value $\alpha_s = 0.30$
is taken.

\begin{figure}
\epsfig{figure=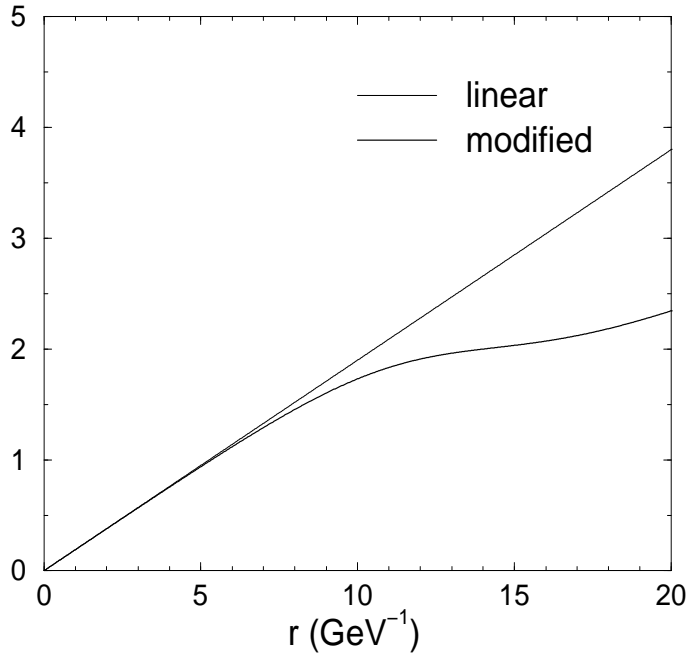,height=90mm,width=90mm}
\caption{Modified potential with parameters given in Eq.~(\ref{eq.43}).
For reference a simple linear potential with $\sigma = 0.19$ is also
plotted.
\label{fig.1}}
\end{figure}

The best description of the meson spectra was obtained for the following
set of the parameters in $\sigma(r)$:
\begin{equation}
 \gamma = 0.40, \; R_1 = 6 \,{\rm GeV}^{-1}, \; B= 20.0,
\label{eq.43}
\end{equation}
while the values of $\sigma_0 = 0.185 \pm 0.005$ GeV${}^2$ and
$\alpha_s = 0.30\pm0.08$ can vary in narrow ranges. In Fig.~\ref{fig.1}
this potential is drawn for the parameters Eq.~(\ref{eq.43}), $\alpha_s=0$
and $\sigma_0= 0.19$ GeV${}^2$.

At the distance $R_2(nL) = 2.5$ fm the value $\sigma_2  = 0.116$
GeV${}^2$ turns out to be rather small, a value 40\% smaller than
$\sigma_0 = 0.19$ GeV${}^2$.  

The spin-averaged meson masses $\bar{M}(nL)$ are calculated solving the
SSE Eq.~(\ref{eq.9}) with the modified potential $V(r)$. Their values
will be given in the next section. Here only some characteristic
features of the SSE solutions for the modified potential will be
discussed.

First, the average values of the string tension $\bar{\sigma} =
<\sigma(r)>_{nL}$ turns out to be almost constant for the ground states
($n_r=0$, $L \leq 3$), while for the radials with $n_r\geq 2$
$\bar{\sigma}$ is already 20\% smaller, see Table~\ref{tab.2}.

\begin{table}
\caption{The values of $\bar{\sigma}(nL)$ (in GeV${}^2$) for the
potential $V(r)$ Eq.~(\ref{eq.40}) with the parameters
Eq.~(\ref{eq.43}) ($\sigma_0 = 0.19$ GeV${}^2$, $\alpha_s = 0.30$).}
\label{tab.2}
\begin{tabular}{|l|r|r|r|r|r|}
\hline
 $n_r\setminus L$ &  0    &    1     &      2   &      3   &      4 \\
\hline 
      0  &    0.188   &   0.187  &   0.184  &   0.179  &   0.168 \\
      1  &    0.183   &   0.176  &   0.164  &   0.153  &   0.150 \\
      2  &    0.165   &   0.155  &   0.154  &   0.154  &   0.150 \\
      3  &    0.156   &   0.156  &   0.154  &   0.154  &   0.150 \\
\hline
\end{tabular}
\end{table}

Secondly, the kinetic energy and therefore the constituent quark mass
$\mu(nL)$ does practically not change for the ground states in contrast
to the high excitations where the string size is growing and the quark
kinetic energy is becoming significantly smaller. In Table~\ref{tab.3}
$\mu(nP)$ for the $nP$ states for the potential Eq.~(\ref{eq.40}) with
$\alpha_s=0$ and the linear potential $\sigma_0 r$ are compared (in
both cases $\sigma_0=0.182$ GeV${}^2$).

\begin{table}
\caption{The constituent masses $\mu(nP)$ for the modified potential
Eq.~(\ref{eq.40}) with $\alpha_s=0$ and $\mu_0(nP)$ for the potential
$\sigma_0r$ with $\sigma_0=0.182$ GeV${}^2$ in both cases.}
\label{tab.3}
\begin{tabular}{|l|r|r|r|r|r|}
\hline
 $n_r$    & 0    &       1       &    2&3         &  4 \\
\hline
 $\mu(nP)$  &    0.424  &    0.430  &      0.441 &      0.559&0.616 \\
 $\mu_0(nP)$ &   0.451  &    0.582  &      0.697 &      0.787&0.872 \\
\hline
\end{tabular}
\end{table}

From Table~\ref{tab.3} one can see that for the $4P (5P)$ states the
difference between the constituent masses is large and reaches $\sim
30$\% . Note that for the modified potential the relations (\ref{eq.12}) are
not valid anymore and
\begin{equation}
 \langle \sigma(r)r \rangle_{nL} >  2\mu(nL)
\label{eq.44}
\end{equation}
and therefore in this case  the string correction $\Delta_{\rm str}$ is given
by a more general expression than in Eq.~(\ref{eq.11}) \cite{ref.7}:
\begin{equation}
 \Delta_{\rm st} = -\frac {\bar{\sigma} <r^{-1}> L(L+1)}
 {\mu(6\mu+ <\sigma(r)r>)}.
\label{eq.45}
\end{equation}
Also in the self-energy term $\Delta_{\rm SE} (nL)$ Eq.~(\ref{eq.14}),
$\bar{\sigma}$ must be used instead of $\sigma_0$.

\section{The masses of the radials} 
\label{sec.5}

The masses of the radial excitations for the modified potential
Eq.~(\ref{eq.40}) with the Coulomb interaction included are presented
in Tables~\ref{tab.4}-\ref{tab.6} for the parameters given in
Eq.~(\ref{eq.43}), $\alpha_s=0.30$, and the string tension $\sigma_0=0.19$
GeV${}^2$. In all cases the calculated meson masses turn out to be in
good agreement with the existing experimental data.

\begin{table}
\caption{The spin-averaged meson masses of the $nS$ states (in GeV) for
the modified potential $V(r)$ Eq.~(\ref{eq.40}) with the parameters 
Eq.~(\ref{eq.43}), $\alpha_s= 0.30$ and $\sigma_0=0.19$ GeV${}^2$ ${}^*$.}
\label{tab.4}
\begin{tabular}{|l|r|r|r|r|r|}
\hline
$nS$   &      $1S$  &      $2S$  &      $3S$  &  $4S$    &   $5S$ \\
\hline
$\bar{M}(nS)$ & 0.618 & 1.400 & 1.868 & 2.176 &  2.502 \\
 &     (0.673)  &   (1.520) &  (2.122)  &  (2.602) &(3.006) \\
 $\bar{M}(nS)_{\rm exp}$ & 0.612 & $1.41\pm 0.02$ & $\pi(1.80)$ & $\rho(2.15)$ & \\
\hline
\end{tabular}

 ${}^*$ The numbers in brackets correspond to the Cornell potential
 Eq.~(\ref{eq.46}) with the same $\alpha_s=0.30$ and $\sigma_0=0.19$
 GeV${}^2$.
\end{table}

The numbers in brackets in Table~\ref{tab.4} are the masses calculated
for the Cornell potential,
\begin{equation}
 V_{\rm C}(r) = -\frac{4\alpha_s} {3r} + \sigma_0 r.
\label{eq.46}
\end{equation}
From a comparison of the numbers given one can see that
with the modified nonperturbative potential the mass of the $4S (5S)$
states appears to be 400 MeV (500 MeV) lower than for the Cornell
potential Eq.~(\ref{eq.46}), while for the $1S$ and $1P$ states the
difference is only about 60 MeV.

We observe the same picture for the $P$-wave and higher excitations
when already for the $2P$ ($3P$) states the spin-averaged mass is
$\simeq 200 MeV$ (350 MeV) smaller due to the modification of the
string potential (see Table~\ref{tab.5}, where the numbers in the
parentheses are calculated with the Cornell potential Eq.~(\ref{eq.46})
with the same  $\alpha_s$ and $\sigma_0$).

\begin{table}
\caption{The spin-averaged meson masses of the $nP$-states (in GeV) for
the potential $V(r)$ Eq.~(\ref{eq.41}) with the parameters
Eq.~(\ref{eq.43}), $\alpha_s = 0.30$ and $\sigma_0=0.19$ GeV${}^2$
${}^*$.}
\label{tab.5}
\begin{tabular}{|l|r|r|r|r|}
\hline
 $nP$ & $1P$ & $2P$ & $ 3P$  & $4P$ \\
\hline
 $\bar{M}(nP)$ & 1.190 & 1.715 & 2.090 & 2.388 \\
	& (1.263) & (1.933) & (2.438) & (2.859) \\
$\bar{M}(nP)_{\rm exp}$ & 1.252 for$\bar{M}(a_J(1P))$ & $a_1(1.70$)
 & $a_1(2.10)$ & $a_1(2.34)$ \\
 & & $a_2(1.75)$ & $a_0(2.05)$ & $f_0(2.34)$  \\
 & 1.245 for $\bar{M}(f_J(1P))$ &$f_2(1.65)$   & $f_0(2.095)$ &\\
\hline
\end{tabular}

${}^*)$ see footnote to Table~\ref{tab.4}.
\end{table}

The calculated masses of the $nS$ and $nP$ states as well as the $nD$
and $nF$ states (see Table~\ref{tab.6}) appear to be in good agreement
with experiment. However, for the $nD$ and $nF$ states a better
agreement is obtained for smaller values of the strong coupling
constant and the numbers given in Table~\ref{tab.6} refer to $\alpha_s=
0.21$. This fact may be connected with a suppression of one-gluon-exchange
for large-size mesons.

\begin{table}
\caption{The spin-averaged meson masses (in GeV) for the $nD$ and $nF$
states for the potential $V(r)$ Eq.~(\ref{eq.41}) with the parameters
Eq.~(\ref{eq.43}), $\alpha_s=0.21$, and  $\sigma_0=0.19$ GeV${}^2$.}
\label{tab.6}
\begin{tabular}{|l|r|r|r|r|r|r|}
\hline
$n$   &   $1D$   &   $2D$   &  $3D$    &   $1F$ & $2F$   &   $3F$ \\
\hline 
 $\bar{M}(nD)$  & 1.628 & 1.973 & 2.290  & 1.926  &2.214 & 2.480 \\
 $M({\rm exp})$ & $pi_2(1.67)$&$\pi_2(2.0)$ & $\pi_(2.25)$ & $a_4(2.01)$
 & $a_4(2.26)$ & - \\
 & $\rho_3(1.69)$ & $\rho_3(1.98)$ & $\rho_3(2.30)$ & $a_3(2.03)$ &
 $a_3(2.28)$ & \\
\hline
\end{tabular}
\end{table}

Thus one can conclude that due to the attenuation of the string
tension in the potential Eq.~(\ref{eq.40}) the masses of the radials
turn out to be $\simeq 100 - 200$ MeV (for the $2L$ states), $\simeq
300$ MeV (for the $3L$ states), and $\simeq 350- 400$ MeV smaller than
for the standard linear potential. It is of interest also to compare
the r.m.s. radii $R(nL)$ of the radials for the modified potential (see
Table~\ref{tab.7}) with that for the $\sigma_0 r$ potential given in
Table~\ref{tab.1}.

\begin{table}
\caption{The r.m.s. radii  $R(nL)$ (in fm) of the $nL$ states for the
potential $V(r)$ (41) with the parameters (43), $\alpha_s=0.30$, and
$\sigma_0=0.19$ GeV${}^2$.}
\label{tab.7}
\begin{tabular}{|l|r|r|r|r|r|}
\hline
  state $\setminus L$ & 0  &  1 & 2  & 3  & 4 \\
\hline
 $1L$  &  0.74  &  1.03  &  1.29  &  1.61 & 2.10 \\
 $2L$  &  1.32  &  1.74  &  2.22  &  2.61 & 2.74 \\
 $3L$  &  2.22  &  2.53  &  2.63  &  2.67 & 2.81 \\
 $4L$  &  2.58  &  2.66  &  2.81  &  2.97 & 3.09 \\
\hline
\end{tabular}
\end{table}

From this comparison one can find out that the size of the $2P$ mesons
is changing from $R(2P) = 1.31$ fm to 1.74 fm while the mass
$\bar{M}(2P)$ is shifted down by about 200 MeV. From Table~\ref{tab.7}
it is also seen that for the modified potential highly excited radials,
like the $3P$ mesons, have very large r.m.s. radii  $\simeq 2.5- 2.8$
fm, in particular the experimentally observed mesons 
$\rho(4S)$ and $a_J(3P)$ have  $R( nL) \sim 2.5- 2.6$ fm.

\section{The slope of the $n_r$-trajectories}
\label{sec.6}

There are not many radial excitations with well established masses
which are included in the PDG compilation \cite{ref.1}. Most radials
were observed in the BNL and Crystal Barrel experiments and discussed
in many papers for the last five years (see
Ref.~\cite{ref.2}-\cite{ref.4} and references therein). Here we present
the values of the slope $\Omega$ defining the $n_r$-trajectory
Eq.~(\ref{eq.1}).

Since we have calculated here only the spin-averaged  masses of the
radials, correspondingly just for them the $n_r$-trajectory
Eq.~(\ref{eq.1}) will be calculated below. Although in many cases there
exists a large uncertainty in the values of $\bar{M}(nL)$ we give below
in Table~\ref{tab.8} several well established masses taking into
account that the spin splittings are small.

\begin{table}
\caption{The experimental spin-averaged masses of the radials}
\label{tab.8}
\begin{tabular}{|l|r|r|r|r|r|r|}
\hline
 $L$ & $1S$ & $2S$ & $1P$ & $2P$ & $3P$ & $4P$ \\
\hline
 $\bar{M}(nL)$ & 0.612 & 1.42$\pm 0.04$ & 1.25$\pm 0.05$ & 1.70$\pm 0.05$
 & 2.07$\pm 0.03$ &  $\sim 2.34$\\
\hline
 $L$ & $     1D $ & $   2D $ & $   3D  $ & $  1F $ & $   2F $ & \\
\hline
 $\bar{M}(nL)$ & 1.67$\pm0.02$ & 2.00$\pm0.02$ &  $\simeq 2.30$ &
 2.02$\pm0.01$ & 2.30$\pm0.01$ & \\
\hline
\end{tabular}
\end{table}

Then taking the difference between the neighbouring $\bar{M}^2$ values 
one can calculate the values of $\Omega$ in Eq.~(\ref{eq.1}):
\begin{eqnarray}
 \Omega_{\rm exp}(S) & = & 1.64 \pm 0.11 \, {\rm GeV}^2, \nonumber \\
 \Omega_{\rm exp}(P) & = & 1.35 \pm 0.25 \, {\rm GeV}^2, \nonumber \\
 \Omega_{\rm exp}(D) & = & 1.21 \pm 0.16 \, {\rm GeV}^2, \nonumber \\
 \Omega_{\rm exp}(F) & = & 1.21 \pm 0.08 \, {\rm GeV}^2,
\label{eq.47}
\end{eqnarray}
which in some cases have a rather large experimental error, but for $L
\neq 0$ practically coincide with the value $\Omega=1.15-1.30$
GeV${}^2$ obtained in Ref.~\cite{ref.3}. Unfortunately, among the $nS$
states the very important ones $\rho(3S)$ and $\pi(4S)$ are still not
observed and the known value of $\Omega_{\rm exp}(nS)$ obtained from the
difference of $\bar{M}^2(2S)$ and $\bar{M}^2(1S)$ appears to be 20\%
larger than in Ref.~\cite{ref.3} and close to our number Eq.~(\ref{eq.49}).

From the meson masses given in Tables~\ref{tab.4}-\ref{tab.6} one can
calculate now the theoretical values of the slope $\Omega_{\rm thr}$
taking for the mass $\bar{M}^2(1L)$  with $n_r=0$ in Eq.~(\ref{eq.1}) the
experimental number. Then one finds
\begin{eqnarray}
 \Omega_{\rm th}(P) & = & 1.38 \pm0.05  \,{\rm GeV}^2, \nonumber \\
\Omega_{\rm th}(D) & = & 1.29 \pm0.06  \,{\rm GeV}^2, \nonumber \\
 \Omega_{\rm th}(F) & = & 1.22 \pm 0.13 \,{\rm GeV}^2.
\label{eq.48}
\end{eqnarray}
However, for the $nS$ radials the slope was found to be 
slightly dependent on $n_r$.
\begin{equation}
\Omega_{\rm th}(S) = 1.60 - 1.45 \; {\rm for}\, n_r=0,1,2,
\label{eq.49}
\end{equation}
which is close to the experimental value $\Omega_{\rm exp} =1.6 \pm
0.1$ GeV${}^2$ obtained for the first excited state, while for higher
$S$-excitations $\Omega_{\rm th}$  was found to be $ \sim 15\%$ smaller
and equal to
\begin{equation}
 \bar{M}^2(4S)- \bar{M}^2(3S) = 2.18^2  -  1.87^2 =  1.26 \,{\rm GeV}^2,
\label{eq.50}
\end{equation}
We can conclude that in the physical picture where the confining
potential is modified due to $q\bar{q}$ pair creation, the slope
$\Omega $ is decreasing from a value $\sim 2.0$ GeV${}^2$ for the
standard linear potential to values in the range $1.2 - 1.35$ GeV${}^2$
for mesons with $L \neq 0$.

\section{Conclusions}
\label{sec.7}
        
We have considered here the light meson orbital and radial excitations
using the effective Hamiltonian derived from QCD under definite and
verifyable assumptions. In the QCD string approach the spin-averaged meson
mass $\bar{M}(nL)$ can be calculated through the only scale parameter -
the string tension and does not contain any arbitrary subtraction
constants, since nonperturbative quark mass renormalization is taken
into account as in Ref.~\cite{ref.12}. The suggested formalism  allows
to resolve three old painstaking problems:

\vspace{3mm}

To determine the origin of the constituent mass for a light quark which
is derived to be the average of the quark kinetic energy operator and
can be computed through the string tension;

\vspace{3mm}

To obtain the correct slope of the Regge trajectory when the string
moment of inertia is taken in account;

\vspace{3mm}

To obtain the correct absolute values of the light meson masses and as
a consequence the correct value of the $L$-trajectory intercept (which
refers to the spin averaged meson masses). The use of the
$L$-trajectories is very convenient since they are universal, i.e. in
the closed-channel approximation they are the same for isovector and
isoscalar mesons.

\vspace{5mm}

In Ref.~\cite{ref.7} this formalism was successfully applied to the
orbital excitations with $n_r=0$ when for the linear confining potential
the string tension was taken constant. However, in an attempt to
describe the radial excitations one encounters a serious problem - the
Regge slope of the $n_r$-trajectories calculated with the same potential
appears to be 1.5-1.7 times larger than in experiment.

This phenomenon, the lowering of  the masses of the highly excited
mesons, is connected in our picture with large sizes of the high
excitations, which can be as large as 2.5 fm and lead to the
formulation of the concept of the predecay (prehadronization) region
where due to $q\bar{q}$ pair creation the string tension is attenuated
at separations $r \geq R_1 = 1.2 fm - 1.4$ fm. In this physical picture
it is important to take into account the specific character of the
$\pi$-meson interaction with a light quark which occurs at the end of
the string and therefore the string does not break due to the
$\pi$-meson emission and this fact reconciles a high probability of
pionic exchanges for the mesons with the existence of linear Regge
trajectories.

The explicit and very simple model of the modified confining potential
where the string tension depends on the separation $r$ for $r \geq
R_1$, allows to obtain the masses of the radials in good agreement with
experiment and may be considered as an explanation for the observation
that the masses of high excitations are lowered. In particular the
centers of gravity of the $2P_J$ and $3P_J$ multiplets appear to be
lower by $\simeq 200$ MeV and $\simeq 350$ MeV respectively. The slope
of the $n_r$-trajectory $\Omega$ ($L\neq 0$) is found to be 1.35- 1.22
GeV${}^2$ in agreement with the analysis in Ref.\cite{ref.3}. For the
$nS$ states the calculated slope is found to be larger, $\Omega \sim
1.5$ GeV${}^2$. The mechanism of reduced string tension has a universal
character and does not depend on the quantum numbers and concrete
positions of open thresholds in meson decays.

\vspace{5mm}
   
The authors would like to express their gratitude to the theory group
of Thomas Jefferson National Accelerator Facility (TJNAF)for their
hospitality. One of the authors (Yu.A.S.) was supported by DOE
contract DE-AC05-84ER40150 under which SURA operates the TJNAF. This
work was partly supported by the RFFI grant 00-02-17836 and INTAS grant
00-00110.

\appendix
\section{Matrix elements for the solutions using the modified
linear potential}
\label{sec.A}

Here we present some characteristics of the SSE solutions
Eq.~(\ref{eq.9}) for the modified confining potential Eq.~(\ref{eq.40})
needed to calculate the spin-averaged meson masses. The eigenvalues of
the Eq.~(\ref{eq.9}) for the potential Eq.~(\ref{eq.40}) plus Coulomb
potential with parameters Eq.~(\ref{eq.43}) and $\sigma_0 = 0.19$
GeV${}^2$ are given in Table~\ref{tab.9} for the $S$- and $P$-wave
mesons ($\alpha_s=0.30$), and in Table~\ref{tab.10} for the $D$- and 
$F$-mesons ($\alpha_s=0.21$).

\begin{table}
\caption{The eigenvalues of the SSE Eq.~(\ref{eq.9}) in GeV for the
Coulomb plus modified confining potential Eq.~(\ref{eq.40}) with the
parameters Eq.~(\ref{eq.43}), $\sigma_0=0.19$ GeV${}^2$ and
$\alpha_s=0.30$ ($L \leq 4,n_r \leq 3$)}
\label{tab.9}
\begin{tabular}{|l|r|r|r|r|r|}
\hline
 $n_r\setminus L$ &    0   &       1   &       2   &      3    &      4 \\
\hline
    0   &      1.204    &  1 721    &   2.098   &   2.402  &  2.645 \\
    1   &      1.858    &  2.190    &   2.436   &   2.628  &  2.799 \\
    2   &      2.268    &  2.471    &   2.656   &   2.840  &  3.013 \\
    3   &      2.542    &  2.730    &   2.904   &   3.065  &  3.218 \\
\hline
\end{tabular}
\end{table}
  
We give also the constituent masses and the matrix elements $<r^{-1}>$
entering the string corrections $\Delta_{\rm str}(nL)$ and $\Delta_{\rm
SE}(nL)$ while the average values of
$\bar{\sigma} = <\sigma(r)>_{nL}$ are given in Table~\ref{tab.2}.

\begin{table}
\caption{The constituent masses (in GeV) and the matrix elements
$<r^{-1}>$ (in GeV${}^{-1}$) for the $P$- and $D$-waves for the same
potential as in Table~\ref{tab.9}.}
\label{tab.10}
\begin{tabular}{|l|r|r|r|r|}
\hline
    $n_r$        &   $1P$    &      $2P$  &       $3P$   &     $4P$ \\
\hline
    $\mu(nP)$    &  0.464    &     0.484  &    0.535     &    0.587 \\
    $<r^{-1}>$  &  0.250    &     0.203  &     0.180    &    0.168 \\
\hline
      &            $1D$    &      $2D$  &      $3D$    &     $4D$ \\
\hline
    $\mu(nD)$   &  0.526    &  0.527   &        0.529   &     0.620 \\
    $<r^{-1}>$ &  0.186    &    0.137  &        0.134  &     0.135 \\
\hline
\end{tabular}
\end{table}

We would like to note that for the modified confining potential the
constituent mass grows by only  about 15\% for the $D$-wave states and
about 25\%  for the $P$-wave states, in contrast to the situation for
the standard linear potential (see Table~\ref{tab.3}) where this growth
is substantially larger.  Also the matrix elements $<r^{-1}>$ for the
$nD$ states with $n_r=1,2,3$ turn out to be equal within 2\%.

\end{document}